\documentclass[10pt,aps,prb,twocolumn,amsmath,amssymb]{revtex4}
\usepackage{amsmath}
\usepackage{graphicx}
\usepackage{color}
\usepackage{multirow}

\begin{document}

 \title{Multireference linearized Coupled Cluster theory for strongly correlated systems using Matrix Product States}
\author{Sandeep Sharma}
\email{sanshar@gmail.com}
\affiliation{Max Planck Institute for Solid State Research, Heisenbergstra{\ss}e 1, 70569 Stuttgart, Germany}
\author{Ali Alavi}
\email{a.alavi@fkf.mpg.de}
\affiliation{Max Planck Institute for Solid State Research, Heisenbergstra{\ss}e 1, 70569 Stuttgart, Germany\\
 Dept of Chemistry, University of Cambridge, Lensfield Road, Cambridge CB2 1EW, United Kingdom}
\begin{abstract}
We propose a multireference linearized coupled cluster theory using matrix product states (MPS-LCC) which provides remarkably
accurate ground-state energies, at a computational cost that has the same scaling as multireference configuration interaction singles and doubles (MRCISD), for a wide variety of electronic Hamiltonians. These 
range from first-row dimers at equilibrium and stretched geometries, to highly multireference systems such as the chromium dimer and lattice models such as periodic two-dimensional 1-band and 3-band Hubbard models.
The MPS-LCC theory shows a speed up of several orders of magnitude over the usual DMRG algorithm while delivering energies in 
excellent agreement with converged DMRG calculations. Also, in all the benchmark calculations presented here MPS-LCC outperformed the commonly used multi-reference quantum chemistry methods in some cases giving energies in excess of an order of magnitude more accurate.   
As a size-extensive method that can treat large active spaces, MPS-LCC opens up the use of multireference quantum chemical techniques in strongly-correlated \emph{ab-initio} 
Hamiltonians, including two and three-dimensional solids. 
\end{abstract}
\maketitle

\section*{Introduction}
One of the most pressing theoretical questions in electronic structure theory is how to deal with realistic strongly correlated electronic systems, 
which typically exhibit combinatorial complexity in the description of the ground-state wavefunction. There are two dominant 
paradigms in electronic structure algorithms, namely {\em variational} and {\em projective} methods, which for different reasons 
struggle to capture the entirety of the problem. Variational methods  (such as CI\cite{Knowles1984} and DMRG\cite{White1992, White1993})  
are in principle robust techniques, but     
generally scale exponentially in the number of correlating orbitals and do not provide size-extensive energies (except in
unreachable exact limits),  whilst projective methods, such as many-body perturbation theory and coupled cluster theory, spectacularly successful in describing 
weak correlation, fail as the underlying single-reference wavefunction upon which they are build diminishes in importance with growing strength of 
correlation. It is natural to ask if a judicious combination of a variational and a projective method exists which can tractably handle 
realistic strong-correlation systems. Here we propose such a method that is based on the
linearized Coupled-Cluster (LCC) method\cite{RevModPhys79} and is implemented using the matrix-product states (MPS)\cite{Schollwock2011} formalism to capture 
highly multi-configurational zeroth- (and higher) order wavefunctions. 

To give a consolidated presentation, we first start by deriving the governing equations for the single reference LCC method which uses the Hartree-Fock wavefunction as the zeroth order state. We show that LCC is closely related to the commonly used variational method, the configuration interaction with singles and doubles (CISD). We highlight the strengths and weaknesses of LCC compared to CISD. The main shortcoming of LCC (it becomes divergent when near-degeneracies are present) can be overcome by using a multireference zeroth order wavefunction. These multireference LCC equations have the same relation to the variational equations usually solved using DMRG\cite{legeza-rev,C0CP01883J,kurashige,White1999,wouters14,Zgid2008}, as the LCC equations have to CISD. We show how the Matrix Product States can be used to efficiently solve the multireference  equations by a small modification of the DMRG algorithm, resulting in a method which we call MPS-LCC. The resulting method is extremely powerful and we demonstrate its strength by solving several tough paradigmatic benchmark problems: first-row dimers at equilibrium and stretched geometries, the 1-band and 3-band (cuprate-like) Hubbard models in the strong-correlation regime $U/t=4-10$ and the Cr$_2$ dimer. 

\section*{Theory}
To derive linearized coupled cluster equations one starts with the coupled cluster singles and doubles wavefunction written using the exponential ansatz $|\Psi\rangle = e^{\hat{T}}|\Phi_0\rangle$, where $\hat{T}=\hat{T}_1+\hat{T}_2$ is the sum of the single and double excitation operators and $|\Phi_0\rangle$ is the Hatree-Fock wavefunction. When the CC wavefunction is substituted into the Schroedinger equation $\hat{H}|\Psi\rangle = E|\Psi\rangle$ and is left multiplied by $e^{-\hat{T}}$ we obtain
 $e^{-\hat{T}}\hat{H}e^{\hat{T}}|\Phi_0\rangle=E|\Phi_0\rangle$. Left projecting onto the Hartree Fock and a set of singly and doubly excited determinants $|\Phi_\mu\rangle$ we obtain the expression for the coupled cluster energy and the governing equations for the $t-$amplitudes.
\begin{align}
E =& \langle\Phi_0|e^{-\hat{T}}\hat{H}e^{\hat{T}}|\Phi_0\rangle  \label{eq:ccE}\\
0 = &\langle\Phi_\mu|e^{-\hat{T}}\hat{H}e^{\hat{T}}|\Phi_0\rangle  \label{eq:ccT}
\end{align}
The set of non-linear Equations~\ref{eq:ccT} can be solved to evaluate the $t-$amplitudes which can then be substituted into Equation~\ref{eq:ccE} to obtain the coupled cluster energy. To obtain LCC equations the above equations can further be simplified by expanding the exponential using the Baker-Campbell-Hausdorff expansion and truncating the series at the first order, i.e $e^{-\hat{T}}\hat{H}e^{\hat{T}} = \hat{H} + [\hat{H}, \hat{T}]$, to yield

\begin{align}
E &=\langle\Phi_0|\hat{H}|\Phi_0\rangle + \langle\Phi_0|\hat{H}|\Psi_1\rangle\label{eq:lccE}\\
0 &=\langle\Phi_\mu|\hat{H}|\Phi_0\rangle + \langle\Phi_\mu|(\hat{H} - E_0)|\Psi_1\rangle  \label{eq:lccT}
\end{align}

Equations~\ref{eq:lccE} and Equations~\ref{eq:lccT} are the governing equations of the single reference linearized coupled cluster theory, where we have defined $|\Psi_1\rangle\equiv\hat{T}|\Phi_0\rangle$ is the LCC correction to the Hartee-Fock wavefunction consisting of only single and double excitations. Equation~\ref{eq:lccT} is now a linear equation in the unknown $|\Psi_1\rangle$ which can be solved and substituted into Equation~\ref{eq:lccE} to calculated the LCC energy.  

Now let us recall that the variational principle can be written as a set of equations $\langle\Phi_\mu|H-E|\Psi\rangle = 0$, where $|\Phi_\mu\rangle$ are the basis states used to expand the wavefunction $|\Psi\rangle$. When the variational principle is used to optimize the CISD wavefunction we obtain 
 \begin{align}
E &=\langle\Phi_0|\hat{H}|\Phi_0\rangle + \langle\Phi_0|\hat{H}|\Psi_1\rangle\label{eq:ci}\\
0&=\langle\Phi_\mu|\hat{H}|\Phi_0\rangle + \langle\Phi_\mu|(\hat{H} - E)|\Psi_1\rangle,  \label{eq:ci}
\end{align}
where we have used intermediate normalization ($\langle\Phi_0|\Psi\rangle = 1$) and as before $|\Psi_1\rangle$ is the correction to the Hartree-Fock wavefunction consisting of only singly and doubly excited determinants. These equations are remarkably similar to Equations~\ref{eq:lccE} and \ref{eq:lccT} with the small modification that the zeroth order energy $E_0$ in Equation~\ref{eq:lccT} is replaced with the variational energy $E$ in Equation~\ref{eq:ci}. This seemingly small change has a significant implication that the LCC energies unlike the CISD energies are fully size-extensive. The LCC energies are nearly equal to the full CCSD energies for weakly correlated problems. We demonstrate this by showing the correlation energy of diamond in Table~\ref{tag:solid} with an \emph{ab-initio} Hamiltonian on a $2\times 2\times2$ $k-$point sampling resulting in 64 electrons in 64 orbital problem (see [\onlinecite{Booth2013}] for more details). 
\begin{table}
\caption{The correlation energy (E$_h$/electron) of diamond calculated using various theories with an \emph{ab-initio} Hamiltonian. The DMRG calculation is still very far from convergence even though it was performed using an MPS with a large virtual bond dimension of $M=7500$. The LCC energies are of comparable accuracy to the CCSD and the FCIQMC energies and were calculated using a small MPS with $M=500$, representing a speed up of over 3 orders of magnitude over the DMRG calculation. (See the main text for algorithmic details.) }\label{tag:solid}
\begin{tabular}{ccccc}
\hline
\hline
FCIQMC& MP2 & CCSD & LCC & DMRG\\
\hline
-0.6190 &-0.5145& -0.6134&  -0.6235	& -0.5477	\\
\hline
\end{tabular}
\end{table}

The size-extensivity unfortunately comes at the cost of variationality and more problematic is the fact that the equations are prone to divergence\cite{Bartlett1981} in cases of near degeneracy when other determinants besides the Hartree-Fock have energies similar to $E_0$. 

This shortcoming can be overcome by formulating a multireference LCC method in which the $E_0$ and $\Psi_0$ in Equation~\ref{eq:lccT} is replaced by the energy and wavefunction obtained by fully correlating a set of orbitals around the Fermi surface. Such a multireference LCC method has been formally derived in the past by Bartlett et al.\cite{Laidig1984,Laidig1987} as well as by Fink\cite{Fink2006, Fink2009}. Here, we use Fink's formulation in which the single reference LCC is written as a perturbation theory using an ingenious use of a zeroth order Hamiltonian. This perturbation theory is then straight forwardly extended to multireferences cases thus resulting in a multireference LCC theory. The advantage of Fink's formulation is that it not only reduces to the LCC equations at the first order but systematic higher order corrections can also be generated. 

Following Fink's formulation we first
start by dividing all the orbitals into an active (or correlating) set, in which 
the orbital occupancies can be 0,1,2, a core set in which the occupancies are constrained to be 2, and a virtual set with zero occupancy. Then, we partition the full Hamiltonian $\hat{H}$, whose ground-state wavefunction $\Psi$ we seek, in terms of number-preserving operators within each subset:
\begin{eqnarray}
\hat{H}&=&\sum_{ij} t_{ij} a^{\dagger}_i a_j + \sum_{ijkl} \langle ij |kl\rangle a^\dagger_i a^\dagger_j a_l a_k = \hat{H}_0 + \hat{U}   \\
\hat{H}_0&=& \sum_{\substack{ij;\\ \Delta n_{ex} = 0}} t_{ij} a_i^{\dag}a_j + \sum_{\substack{ijkl;\\ \Delta n_{ex} = 0}} \langle ij |kl\rangle a_i^{\dag}a_j^{\dag}a_la_k \label{eq:part} 
\end{eqnarray}
The constraint $\Delta n_{ex}=0$ implies that the operators do not transfer electrons between the three sets of orbitals, and $\hat{U}$ 
contains all remaining terms.  

To begin with we will assume   
that the ground-state eigenfunction $\Phi^{(0)}$ of $\hat{H}_0$ can be found (later this assumption will be relaxed using the \emph{projector} approximation). 
This zeroth order wavefunction (which will in general have a combinatorial
complexity) has an  eigenvalue equal to the expectation value of the full Hamiltonian $E^{(0)} = \langle\Phi^{(0)}|H|\Phi^{(0)}\rangle$ and the first order energy is zero. It is possible to develop the usual perturbation theory master equations to express successive corrections  ($\Phi^{(n)}$), to the wavefunction: $\Psi=\Phi^{(0)}+\Phi^{(1)}+\Phi^{(2)}+...$. We expect this series to converge if the norms at each subsequent
order rapidly diminish. In the limit in which the active space spans all orbitals, we recover full CI (which is always convergent), 
whereas in the opposite limit where there are
zero orbitals in the active space, we recover linearized Coupled Cluster theory, itself an excellent weak-correlation theory, but which diverges in strong correlation systems. We expect to have convergent theory for any level of correlation as long as
the active space is sufficiently large. 

The equations governing $\Phi^{(n)}$ are shown in Equation~\ref{eq:1}, where $P$ is the projector on to the zeroth order wavefunction ($P=|\Phi^{(0)}\rangle\langle\Phi^{(0)}|$) and $Q$ is its complement ($1-P$); the set of linear equations must be solved one at a time to obtain the $n^{th}$ order correction to the wavefunction ($\Phi^{(n)}$)\cite{Helgaker2000}. Once  $\Phi^{(n)}$ is known, $2n$ and $2n+1$ order corrections to the energy ($E^{2n}$, $E^{2n+1}$) can be calculated due to Wigners $(2n+1)$ rule using Equation~\ref{eq:2}. 
\begin{align}
(\hat{H}_0 - E^{(0)}) |\Phi^{(n)}\rangle = Q\left(-\hat{U} |\Phi^{(n-1)}\rangle + \sum_{k=1}^{n} E^{(k)} |\Phi^{(n-k)}\rangle\right) \label{eq:1}\\
E^{(2n)}  = \langle \Phi^{(n-1)}| \hat{U} |\Phi^{(n)}\rangle - \sum_{k=1}^{n}\sum_{j=1}^{n-1} E^{(2n-k-j)} \langle\Phi^{(k)}|\Phi^{(j)}\rangle \nonumber\\
E^{(2n+1)}  = \langle \Phi^{(n)}| \hat{U} |\Phi^{(n)}\rangle - \sum_{k=1}^{n}\sum_{j=1}^{n} E^{(2n+1-k-j)} \langle\Phi^{(k)}|\Phi^{(j)}\rangle \label{eq:2}
\end{align}

We use the variational principle for the perturbation theory\cite{oktay,hylleraas} which states that to solve Equation~\ref{eq:1} it is sufficient to minimize the Hylleraas functional shown in Equation~\ref{eq:1b} with respect to $|\Phi^{(n)}\rangle$. Only a small modification to the DMRG algorithm, implemented in the \textsc{Block} code\cite{Chan2002, sharmaspin, Sharma2015}, was required to minimize the Hylleraas functional. The details of the algorithm are outlined in the next section\cite{Sharma2014}. 

\emph{Projector approximation:} The cost of optimizing the zeroth order wavefunction scales exponentially with the size of the active space. This exponential cost can be circumvented by only approximately diagonalizing the zeroth order hamiltonian. The error in the zeroth order wavefunction can then also be corrected perturbatively. To do this we define a new zeroth-order hamiltonian $\tilde{H}_0 = P\hat{H}_0P + Q\hat{H}_0Q$, and the perturbing hamiltonian becomes $\tilde{U} = P\hat{H}_0Q + Q\hat{H}_0P + \hat{U}$. It can then be shown that the use of the modified zeroth order hamiltonian changes the Equations~\ref{eq:1} only at the first order to $(\hat{H}_0 - E^{(0)}) |\Phi^{(1)}\rangle = Q\hat{H} |\Phi^{(0)}\rangle$, where $\hat{H}$ is the unperturbed hamiltonian; and the expression of the second order energy changes from $\langle \Phi^{(0)}| \hat{U} |\Phi^{(1)}\rangle$ to $\langle \Phi^{(0)}| \hat{H} |\Phi^{(1)}\rangle $.
These results are remarkable because in essence they imply that even approximate solution of the zeroth order equation is sufficient and the original perturbation series can be used with only minor modifications.\\

Before moving on to describe our implementation, we
would like to point out that Fink's formulation is different than Bartlett's formulation of LMRCC theory. Unlike Bartlett's equations, in Fink's equations the states with different number of electrons in the active, core and virtual spaces don't interact with each other through the zeroth order Hamiltonian. Also, unlike Bartlett's equations the present LMRCC equations cannot be obtained straightforwardly from governing equations of ic-MRCC\cite{Evangelista2012, Hanauer2011} by discarding terms that are higher order than linear in the
excitation operator $T$. A key difference between the current approach and the commonly used approach in multi reference methods is that we use a single MPS to represent the first order wavefunction instead of expanding it in the space formed by internally contracted singles and doubles excitations out of the reference wavefunction. The states forming this space are not mutually orthogonal and without special care the equations can become ill-conditioned due to linear dependencies. The problem of linear-dependencies never arise in our formulation, further unlike the usual formulations at convergence our wavefunction is allowed to relax in the full uncontracted space of singles and doubles excitations. Another important difference is that unlike the usual formulation we do not need any reduced density matrices, for example up to sixth-order RDMs are required for a naive implementation of ic-MRCI calculations, although in practice this requirement can be realxed by using configuration state functions instead of internally contracted states\cite{Shamasundar2011, Werner1988}. The two other approaches used for incorporating post-DMRG dynamical correlation are the Canonical Transformation theory and the methods based on exploring the tangent space of the MPS\cite{Haegeman2011,Wouters2013,Nakatani2014}. The canonical transformation theory\cite{dmrgct, Neuscamman2010} tries to perform the unitary multi reference coupled cluster theory, but with the simplification that all RDMs higher than the two-body RDMs are evaluated using the cumulant approximation. The tangent space based methods are in principle quite similar to our current method with the main difference that the corrections to the reference MPS are restricted to linear combination of its tangent space vectors. This is a far more restrictive space than the one used here which is the one spanned by a single MPS with an arbitrarily large bond dimension.

We would also like to emphasize that this method can be extended in several ways. First, the static one-particle and two-particle correlation functions of the ground state can be easily calculated\cite{Ghosh2008, Zgid20082bdm}.  Second, in addition to the ground state, a set of low-lying excited states can also be calculated with a computational cost that scales linearly with the number of states using quasi-degenerate perturbation theory\cite{Klein1974,Shavitt1980,brandow}. Finally, dynamical correlation functions can be calculated by combining the working equations of coupled cluster Green's function framework\cite{CCGF} with the dynamical DMRG\cite{DDMRG-white,DDMRG}.\\

\section*{Implementation}
\begin{figure}[h]
\begin{center}
\includegraphics[width=0.3\textwidth]{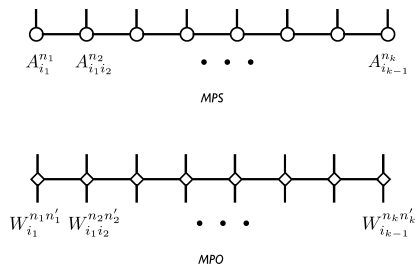}
\end{center}
\caption{ A matrix product state (MPS) can be represented graphically using a series of 3-dimensional tensors in which, 
the physical index (pointing upwards) of the tensors denotes the occupation of the orbital and the other two indices, known as virtual indices, are 
sequentially contracted.  
Similarly, a matrix product operator (MPO) can be represented graphically using a series of 4-dimensional tensors, with two physical indices and two virtual indices. The virtual indices of adjacent tensors are contracted sequentially.}
\label{fig:mpsmpo}
\end{figure}

\begin{figure}
\begin{center}
\includegraphics[width=0.2\textwidth]{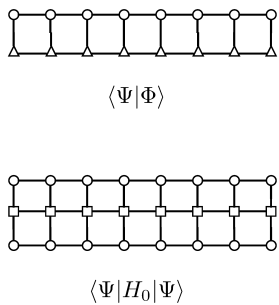}
\end{center}
\caption{ The figure shown the overlap and a transition matrix element of an MPO $H_0$ between MPS $\Psi$ and $\Phi$. These are calculated by contracting the free physical dimension of the MPS and MPO sequentially as shown in the figure. By appropriately ordering the sequence of these contractions it can be shown that the cost of evaluating an overlap is $O(kM^3)$ and a transition matrix element is $O(k^3M^3)$ respectively. }
\label{fig:ops}
\end{figure}

A general MPS representing a wavefunction $|\Psi\rangle$  is shown in Eq.~(\ref{eq:mps1}), where $n_i$ is the occupation of orbital $i$, and $i_1 \cdots i_{k-1}$ are the virtual indices that are contracted to obtain the final state. By increasing the size of the virtual indices an MPS can be used to represent any wavefunction arbitrarily accurately. Similarly, operator $\Omega$ can be written in a matrix product operator (MPO) form as shown in Eq.~(\ref{eq:mpo}). It should be noted that the expression in Equation~\ref{eq:mpo} is very general, but it can be simplified if one limits the operator to have at most two body interactions. In such cases it can be shown that the virtual bond dimension of the operator need never be greater than $k^2$, where $k$ is the number of orbitals. The algebraic notation very quickly becomes cumbersome due to the rapid proliferation of indices; instead the graphical notation which is briefly explained below is much more convenient and intuitive. Here we can only give a short introduction to the notation, for more details we refer the reader to the excellent review article by Schollw\"{o}ck\cite{Schollwock2011}.

Both the MPS and MPO can be conveniently represented graphically as shown in Figure~\ref{fig:mpsmpo}. In the figures each matrix (strictly speaking this is a three dimensional tensor) of the MPS is represented by a circle with three bonds jutting out, one pointing in the upward direction which corresponds to the physical index $n_i$ and two others pointing horizontally that correspond to the virtual indices. The bonds corresponding to the virtual indices of the adjacent matrices are joined together, which algebraically corresponds to contracting the virtual indices, to obtain the wavefunction. The different MPS and MPO when written graphically are distinguished by the symbols used to represent their matrices; e.g. here we use a circle of $\Psi$, a triangle for $\Phi$ and a square for $H_0$ respectively.

\begin{figure}
\begin{center}
\includegraphics[width=0.3\textwidth]{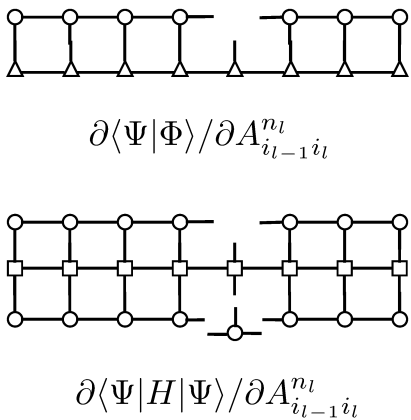}
\end{center}
\caption{ The figure shows the partial derivative of the overlap and transition matrix element with respect to the local tensor $A^{n_l}_{i_{l-1} i_{l}}$ of the MPS. Each of these graphs represents $4M^2$ terms corresponding to taking the partial derivative with respect to each element of the tensor $A^{n_l}_{i_{l-1} i_{l}}$ . }
\label{fig:solve}
\end{figure}

It can be shown that taking the overlap between two MPS and calculating the matrix element of an MPO between two MPS can be performed with a polynomial cpu cost of $O(kM^3)$ and $O(k^3M^3)$ respectively (see Figure~\ref{fig:ops}). To get this computational scaling one needs to perform the various tensor contractions in a specific well-defined order. A suboptimal order of contractions can lead to computational cost that scales exponentially with the number of orbitals $k$. The partial derivative of overlap or operator expectation value with respect to one of the matrices of an MPS gives rise to a tensor which has exactly the same dimension as that of the matrix. This partial derivative in graphical language is represented by graph of the overlap or the expectation with the corresponding matrix removed from it as shown in Figure~\ref{fig:solve}.

In both MPS-LCC and DMRG the functional being optimized is quadratic in the wavefunction of interest. In the case of DMRG it is the energy functional,
\begin{align}
E[\Psi] =& \langle\Phi | \hat{H}| \Phi \rangle - \langle\Phi |E | \Phi \rangle \label{eq:EH}
\end{align}
and in the case of LCC it is the Hylleraas functional shown in Equation~\ref{eq:1b}.
\begin{widetext}
\begin{align}
 |\Psi\rangle =& \sum_{\{n\}, i_1\cdots i_{k-1}} A^{n_1}_{i_1} A^{n_2}_{i_1 i_2} \ldots A^{n_k}_{i_{k-1}} |n_1 n_2 \ldots n_k\rangle. \label{eq:mps1}\\
 \Omega =& \sum_{\{n\}\{n'\}, i_1\cdots i_{k-1}} W^{n_1 n_1'}_{i_1} W^{n_2 n_2'}_{i_1 i_2} \ldots W^{n_k n_k'}_{i_{k-1}} |n_1 n_2 \ldots n_k\rangle \langle n_1' n_2' \ldots n_k'|\label{eq:mpo}\\
 H[\Phi^{(n)}] =& \langle \Phi^{(n)} | \hat{H}_0 - E^{(0)} | \Phi^{(n)} \rangle -  \left( \langle\Phi^{(n)} |\hat{U} |\Phi^{(n-1)}\rangle - \sum_{k=1}^{n} E^{(k)}  \langle \Phi^{(n)} |\Phi^{(n-k)}\rangle \right) &\nonumber\\
&+ \langle \Phi^{(n)} |\Phi^{(0)}\rangle \left( \langle \Phi^{(0)} |\hat{U} |\Phi^{(n-1)}\rangle - \sum_{k=1}^{n} E^{(k)}  \langle \Phi^{(0)} |\Phi^{(n-k)}\rangle \right) \label{eq:1b}
\end{align}
\end{widetext}

In MPS-LCC the wavefunction of interest is written as an MPS and is then evaluated by minimizing the Hylleraas functionals using the sweep algorithm. The key element of the sweep algorithm is that at each sweep iteration $l$ only one tensor $A^{n_l}_{i_{l-1} i_{l}}$ is optimized keeping all the others constant. Figure~\ref{fig:solve} shows the partial derivative of $\langle\Psi|\Phi\rangle$ and $\langle\Psi|H|\Phi\rangle$ with respect to the unknown tensor $A^{n_l}_{i_{l-1} i_{l}}$ of wavefunction $\Psi$. The governing equations that need to be solved at each sweep iteration are obtained by taking the partial derivatives of Equations~\ref{eq:1b} and equating them to zero. This converts a complicated multi-linear problem into a linear algebra problem (a linear equation) with the elements of tensor $A^{n_l}_{i_{l-1} i_{l}}$ as unknowns. Standard iterative algorithms like the, Jacobi-Davidson and Conjugate-Gradient methods can be used to solve the linear algebra problems. By increasing the virtual bond dimension of the MPS expressing $|\Phi^{(n)}\rangle$, the Equation~\ref{eq:1b} can be minimized arbitrarily accurately. The cpu cost per sweep iteration for calculating the two most expensive terms $|\langle\Phi^{(n)} | \hat{H}_0 - E^{(0)} | \Phi^{(n)} \rangle$ and $\langle\Phi^{(n)} | \hat{U}| \Phi^{(n-1)} \rangle$ on the right hand size of Equation~\ref{eq:1b} are $O(k^2M_n^3)$ and $O(k^2M_n^2M_{n-1}) + O(k^2M_nM_{n-1}^2)$ respectively, where $M_n$ is the virtual bond dimension of the MPS representing the state $| \Phi^{(n)} \rangle$. The entire algorithm is implemented in the \textsc{Block} code which includes the ability to treat several different symmetries including $SU(2)$ and non-Abelian point group.

\section*{Benchmarks}
Earlier we showed that MPS-LCC is more efficient that the variational DMRG algorithm for weakly correlated systems like the diamond crystal. Here we demonstrate that it shows equally impressive performances for: first row dimers at equilibrium and stretched geometries, strongly correlated systems like 2-dimensional 1-band and 3-band Hubbard model at half filling and the Cr$_2$ dimer. The half filled 1-band Hubbard model is chosen because reliable AFQMC results are available due to absence of sign-problem. It should be pointed out that from the perspective of MPS-LCC calculations, half-filling represents the hardest case and we expect the quality of results to be better away from half filling. 

\subsubsection*{First row dimers}
\begin{figure}
\begin{center}
\includegraphics[width=0.5\textwidth]{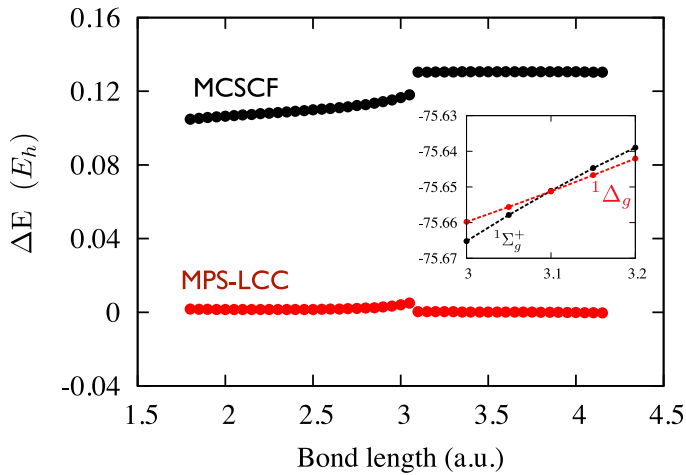}
\end{center}
\caption{The figure shows the error in the energies calculated using MCSCF and MPS-LCC methods relative to the FCI energies for the carbon dimer at various bond lengths using the cc-pVDZ basis set. There is a discontinuity in the energies of the MPS-LCC and MCSCF methods at 3.1 bohr because of a curve crossing between $^1\Sigma_g^+$ and $^1\Delta_g$  states. This curve crossing is shown in the inset.\label{fig:c2curve}}
\end{figure}

We start by calculating the energy of the ground state of the C$_2$ dimer at various bond lengths using MPS-LCC with the double-zeta basis set. The MPS-LCC calculations are performed on an MCSCF reference wavefunction with an active space of (8o, 8e) and the resulting energies are tabulated in Table~\ref{tab:c2bond}. The table also shows the FCI energies which are calculated by correlating all 12 electrons in 28 orbitals using the DMRG algorithm as implemented in the \textsc{Block} code.The errors in the MCSCF and MPS-LCC energies, calculated relative to the FCI energies, are plotted in Figure~\ref{fig:c2curve}. We see that there is a discontinuity in the MCSCF and MPS-LCC energies at a bond length of 3.10 bohr. This is because the $^1\Sigma_g^+$ and $^1\Delta_g$ energy curves cross between 3.05 bohr and 3.10 bohr bond lengths, with the former being the ground state at shorter bond lengths and the latter being the ground state at larger bond lengths. In our calculations the discontinuity in the MCSCF and MPS-LCC curves arise because we have only used the $D_{2h}$ subgroup of the full $D_{\infty h}$ point group of the molecule. Besides the discontinuity at the curve crossing the MPS-LCC energy is both continuous and smooth despite the  fact that at bond lengths greater than 3.05 bohr the ground state and the first excited states are nearly degenerate with a maximum separation of less than 6 mE$_h$.

\addtolength{\tabcolsep}{6pt}
\begin{table} 
\caption{The table shows the ground state energy of the C$_2$ dimer calculated using various methods with the cc-pVDZ basis set. The MCSCF and MPS-LCC use an (8e, 8o) active space and the FCI energy is calculated by fully correlating 12 electrons in 28 orbitals.}\label{tab:c2bond}
\begin{tabular} {l c c c}
\hline
\hline
$r/a_0$ &\multicolumn{3}{c}{Energy/E$_h$}\\
& FCI & MCSCF & MPS-LCC\\
\hline
1.80&-75.4549&-75.3501&-75.4532\\
1.85&-75.5132&-75.4080&-75.5116\\
1.90&-75.5621&-75.4564&-75.5605\\
1.95&-75.6026&-75.4965&-75.6010\\
2.00&-75.6358&-75.5294&-75.6343\\
2.05&-75.6628&-75.5560&-75.6613\\
2.10&-75.6843&-75.5771&-75.6828\\
2.15&-75.7010&-75.5935&-75.6996\\
2.20&-75.7136&-75.6058&-75.7122\\
2.25&-75.7227&-75.6145&-75.7213\\
2.30&-75.7287&-75.6202&-75.7272\\
2.35&-75.7320&-75.6232&-75.7306\\
2.40&-75.7332&-75.6240&-75.7317\\
2.45&-75.7324&-75.6229&-75.7309\\
2.50&-75.7300&-75.6202&-75.7285\\
2.55&-75.7263&-75.6161&-75.7247\\
2.60&-75.7215&-75.6109&-75.7199\\
2.65&-75.7159&-75.6047&-75.7141\\
2.70&-75.7095&-75.5979&-75.7076\\
2.75&-75.7026&-75.5904&-75.7006\\
2.80&-75.6953&-75.5825&-75.6931\\
2.85&-75.6878&-75.5743&-75.6853\\
2.90&-75.6802&-75.5659&-75.6773\\
2.95&-75.6726&-75.5573&-75.6693\\
3.00&-75.6652&-75.5487&-75.6612\\
3.05&-75.6580&-75.5400&-75.6532\\
3.10&-75.6512&-75.5208&-75.6509\\
3.15&-75.6467&-75.5163&-75.6464\\
3.20&-75.6420&-75.5116&-75.6418\\
3.25&-75.6373&-75.5068&-75.6371\\
3.30&-75.6326&-75.5020&-75.6323\\
3.35&-75.6278&-75.4972&-75.6276\\
3.40&-75.6230&-75.4924&-75.6228\\
3.45&-75.6183&-75.4877&-75.6182\\
3.50&-75.6137&-75.4831&-75.6135\\
3.55&-75.6091&-75.4785&-75.6090\\
3.60&-75.6047&-75.4740&-75.6046\\
3.65&-75.6003&-75.4696&-75.6002\\
3.70&-75.5961&-75.4654&-75.5960\\
3.75&-75.5920&-75.4613&-75.5920\\
3.80&-75.5880&-75.4573&-75.5880\\
3.85&-75.5842&-75.4535&-75.5842\\
3.90&-75.5805&-75.4498&-75.5805\\
3.95&-75.5769&-75.4463&-75.5770\\
4.00&-75.5735&-75.4429&-75.5737\\
4.05&-75.5703&-75.4398&-75.5705\\
4.10&-75.5672&-75.4367&-75.5674\\
4.15&-75.5642&-75.4339&-75.5645\\
\hline
\end{tabular}
\end{table}
\addtolength{\tabcolsep}{-6pt}

\begin{figure}
\begin{center}
\includegraphics[width=0.4\textwidth]{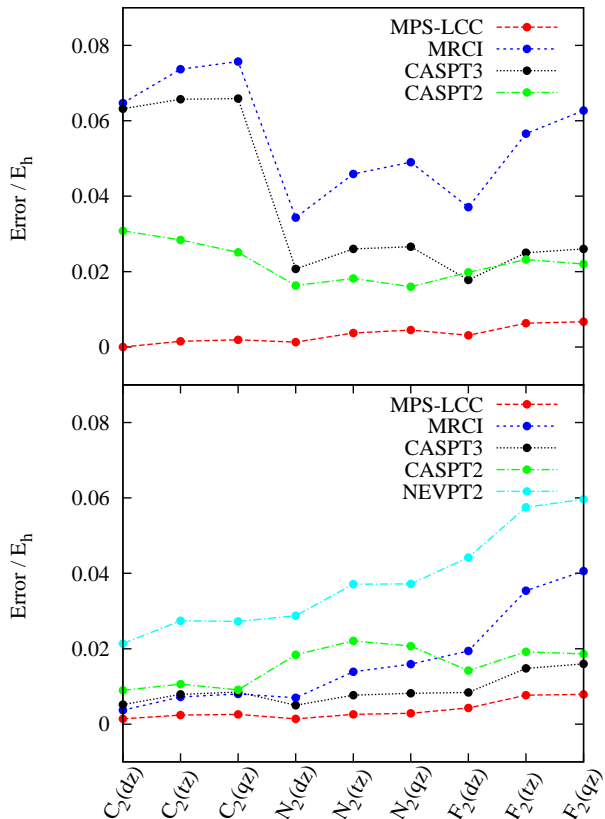}
\end{center}
\caption{The error in the energies calculated using various active space methods relative to the highly accurate FCIQMC energies\cite{Cleland2012} of C$_2$, N$_2$ and F$_2$ molecules at bond lengths of 1.24253, 1.0977 and 1.4119 \AA~respectively. The upper panel shows the errors when the CAS-CI wavefunction was the reference and the bottom panel used an MCSCF wavefunction as reference. In both cases the active space chosen was the full valence space containing eight orbitals.\label{fig:diatoms}}
\end{figure}

We also benchmark the MPS-LCC method for C$_2$, N$_2$ and F$_2$ molecules at their equilibrium bond lengths of 1.24253, 1.0977 and 1.4119 \AA~respectively against the full configuration interaction (FCI) energies calculated using the FCIQMC method with up to quadruple-zeta basis set\cite{Cleland2012}. Here, the energies are calculated using various commonly used active-space methods like MRCI\cite{Knowles1988, Werner1988}, CASPT2, CASPT3\cite{Werner96}, NEVPT2\cite{Angeli2001,nevpt2} and the MPS-LCC method developed in this work. For all these methods two sets of calculations were performed, the first in which a complete active space configuration interaction (CAS-CI) wavefunction was used as a reference and in the second multi-configuration self consistent field (MCSCF) wavefunction was used. Both CAS-CI and MCSCF calculations were performed with frozen core and all the valence orbitals included in the active space. The results of the calculations are shown in Table~\ref{tab:diatoms} and are plotted in Figure~\ref{fig:diatoms}. These calculations show that MPS-LCC gives higher accuracy for these molecules compared to all other methods. One striking feature of these results is the fact that the quality of the MPS-LCC results are almost unaffected by the reference wavefunction. This is a well known feature of the CCSD method, but from these results it looks like the linearized version of the theory also shows this feature as long as the active space is large enough to avoid divergences. 

It can be seen that the results of the MRCI calculation for the F$_2$ dimer are much less accurate than the C$_2$ and N$_2$ dimers. This is most likely due to the relatively small size of the zeroth order wavefunction which only has about 64 determinants in the active space compared to 4900 and 3136 determinants respectively for C$_2$ and N$_2$ respectively. The perturbation theories are in general somewhat less sensitive to the size of the Hilbert space of the zeroth order wavefunction because they are not variational. 

We have also performed calculations on the C$_2$ dimer with cc-pVQZ basis set at various bond lengths. Here the accurate benchmark data is obtained using the frozen core DMRG calculations\cite{Sharma2015} published recently. These results again show the higher accuracy obtained by the MPS-LCC method relative to other methods. There are larger non-parallality errors at bond length of 1.6 \AA~possibly because of curve crosing between two $A_g$ states (the intersecting $^1\Sigma_g^+$ and $^1\Delta_g$ states belong to the $A_g$ irreducible representation in the D$_{2h}$ subgroup) near this geometry. Quasi-degenerate perturbation theory can in principle be used to ameliorate these problems.

\begin{table*}
 \caption{The third column show the ground state energy and estimated uncertainty  in Hartrees ($E_h$) of the C$_2$, N$_2$ and F$_2$ molecules at bond lengths of 1.24253, 1.0977 and 1.4119 \AA~respectively calculated using the FCIQMC\cite{Cleland2012} method. The rest of the table shows the errors of various active space methods relative to FCIQMC in milli-Hartrees ($mE_h$). The active space used for these calculations consisted of the eight valence orbitals including $2s$ and $2p$ orbitals. Two sets of calculations were performed, one with the CAS-CI reference and the other with MCSCF reference.}\label{tab:diatoms}
 \begin{center}
  \begin{tabular}{l c c c c c c c c c c c c c}
  \hline
  \hline
 Molecule &Basis&FCIQMC&&\multicolumn{4}{c}{CAS-CI reference ($mE_h$)}&&\multicolumn{5}{c}{MCSCF reference ($mE_h$)}\\
\cline{5-8}\cline{10-14}
 && ($E_h$)&& MPS-LCC& CASPT2& CASPT3& MRCI&& MPS-LCC& CASPT2& CASPT3& MRCI& NEVPT2\\
 \hline
C$_2$& dz& -75.7285(1)&& 0.0& 64.7& 30.8& 63.2&& 1.4& 9.0& 5.2& 3.7& 21.4\\
C$_2$& tz& -75.7850(1)&& 1.5& 73.7& 28.4& 65.7&& 2.4& 10.6& 7.9& 7.2& 27.4\\
C$_2$& qz& -75.8023(3)&& 1.9& 75.7& 25.1& 65.9&& 2.6& 9.1& 8.5& 8.0& 27.2\\
N$_2$& dz& -109.2767(1)&& 1.3& 34.3& 16.3& 20.7&& 1.4& 18.4& 5.0& 7.0& 28.7\\
N$_2$& tz& -109.3754(1)&& 3.7& 45.9& 18.2& 26.0&& 2.6& 22.1& 7.7& 13.9& 37.1\\
N$_2$& qz& -109.4058(1)&& 4.5& 49.0& 16.0& 26.6&& 2.9& 20.7& 8.2& 15.9& 37.2\\
F$_2$& dz& -199.0994(1)&& 3.1& 37.1& 19.8& 17.8&& 4.3& 14.2& 8.4& 19.4& 44.2\\
F$_2$& tz& -199.2977(1)&& 6.3& 56.6& 23.2& 25.0&& 7.7& 19.2& 14.8& 35.4& 57.5\\
F$_2$& qz& -199.3598(2)&& 6.7& 62.7& 22.0& 26.0&& 7.9& 18.6& 16.0& 40.6& 59.6\\
\hline
  \end{tabular}
 \end{center}
\end{table*}

\begin{figure}
\begin{center}
\includegraphics[width=0.45\textwidth]{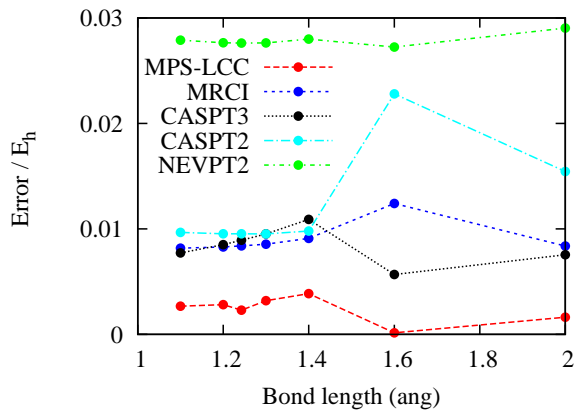}
\end{center}
\caption{The error in the energies of the Carbon dimer using various active space methods relative to the energies calculated using DMRG which themselves have an error of less than 0.1 mH relative to the FCI energies\cite{Sharma2015}. \label{fig:c2}}
\end{figure}

\begin{table}
 \caption{The second row of the table shows the DMRG energy in Hartrees for Carbon dimer at various bond lengths. The rest of the table shows the error in milli-Hartree (m$E_h$)of various methods relative to the DMRG energies. All calculations besides DMRG were performed on a reference wavefunction obtained by a frozen core MCSCF calculation with eight electrons in eight orbitals active space.}\label{tab:c2}
 \begin{center}
  \begin{tabular}{lccccccc}
   \hline
   r&DMRG&&MPS-LCC&MRCI&CASPT3&CASPT2&NEVPT2\\
   \AA&($E_h$)&&\multicolumn{5}{c}{(m$E_h$)}\\
   \cline{1-2}\cline{4-8}
   1.1&-75.7613&& 2.7& 8.2& 7.7& 9.7& 27.9\\
1.2& -75.7992&& 2.8& 8.3& 8.5& 9.5& 27.7\\
1.24253& -75.8027&& 2.3& 8.4& 8.9& 9.5& 27.6\\
1.3& -75.7994&& 3.2& 8.5& 9.5& 9.5& 27.6\\
1.4& -75.7797&& 3.9& 9.1& 10.9& 9.8& 28.0\\
1.6& -75.7241&& 0.1& 12.4& 5.7& 22.8& 27.2\\
2& -75.6460&& 1.6& 8.4& 7.6& 15.4& 29.0\\
\hline
  \end{tabular}
 \end{center}
\end{table}

\subsubsection*{Cr2 dimer} 
The chromium dimer has been a challenging problem for quantum chemistry and large active spaces and basis sets are required to obtain the correct binding curve\cite{Purwanto2015,limanni13,Kurashige2011,Zgid2009,thomas09,Angeli2006,celani04,holger99,Goodgame1985,Andersson1994,Bauschlicher1994}. Here, we do not try to calculate the best binding curve that we can, but instead use some smaller benchmark calculations to compare the commonly used quantum chemical methods against the MPS-LCC method. In particular, we carry out an all electron (48e, 42o) calculation on the Cr$_2$ dimer with an SVP basis set at a bond length of 1.5 \AA. 

Table~\ref{tab:cr2} shows the total energy and the well depth calculated using various quantum chemical methods, including coupled cluster with up to fourth order excitation and the contracted MRCI method\cite{Shamasundar2011} as implemented in \textsc{Molpro}\cite{Werner12}. All the multireference calculations including the LCC were performed with a zero order wavefunction obtained by performing a (12e,12o) CASSCF calculation. The first order MPS-LCC wavefunction was represented with an MPS of virtual bond dimension 4000. It can be seen that the MPS-LCC method not only gives a total energy closest to our best guess of FCI energy but the error in the calculated well depth is over an order of magnitude more accurate than any of the other methods shown here.

\begin{table}
\caption{The table presents the absolute energies and the well depths (in E$_h$) calculated for the chromium dimer at 1.5~\AA ~bond length with an SVP basis set. CASSCF, MRCI and MPS-LCC theories used an active space of 12 electrons in 12 orbitals. Notice that the well-depth calculated using the MPS-LCC method is over an order of magnitude more accurate than the result of any other method presented. The ``FCI" energy was calculated by extrapolating a large DMRG calculation\cite{Olivares2015} to zero discarded weight limit and is estimated to have a residual error of about 2 mE$_h$.}\label{tab:cr2}
\begin{center}
\begin{tabular}{lcc}
\hline
\hline
Method&\multicolumn{2}{c}	{1.5\AA (SVP)}\\			
\cline{2-3}
&Energy& well depth \\
\hline
CASSCF&	-2,086.2256	&0.167\\
CCSD	&-2,086.3880	&0.177\\
CCSD(T)	&-2,086.4222	&0.150\\
CCSDTQ&	-2,086.4302&	0.143\\
MRCIC&	-2,086.4280	&0.138\\
MPS-LCC&	-2,086.4349	&0.129\\
\hline
\bf{FCI}&	\bf{-2,086.4448 $\pm$ 0.002}	&  \bf{0.129}\\
\hline
\end{tabular}
\end{center}
\end{table}

\subsubsection*{2d Hubbard model}
\begin{figure}
\begin{center}
\includegraphics[width=0.4\textwidth]{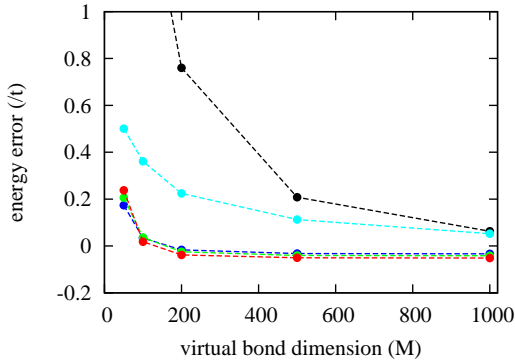}
\end{center}
\caption{The graph shows the energy error of three \emph{p-}MPS-LCC calculations, (where the approximate zeroth order wavefunction is represented with an MPS of bond dimension 100 (blue), 20 (green) and 10 (red) respectively) relative to the FCI energy versus the MPS bond dimension of the first order wavefunction. We have also shown the energy error of the DMRG calculation using localized orbitals (black) and delocalized k-space orbitals (cyan) respectively. The MPS-LCC method shows several orders of magnitude speed up over the corresponding DMRG calculations.\label{fig:discard}}
\end{figure}

We first calculate the ground state energy of a half filled 18-site 2D Hubbard lattice at $U/t=4.0$. In this system when 8 orbitals (4 degenerate orbitals above and 4 below the Fermi surface in $k-$space) are treated variationally and the rest of the orbitals are correlated using the MPS-LCC framework we get excellent agreement with the AFQMC results which are expected to agree with the FCI results to all shown significant digits. The zeroth order wavefunction calculated variationally only captures about 14\% of the correlation energy, but with the inclusion of the first order correction to the wavefunction we account for 101\% of the remaining correlation energy. Even though we don't necessarily recommend performing higher order perturbation theory, for illustration purposes we show that subsequent higher order corrections calculated up to the 7$^{th}$ order show rapid converge towards the FCI energy as shown in Table~\ref{tab:hub}. To access the cost of the method, we show in Figure~\ref{fig:discard} that the third order MPS-LCC energy rapidly converges to its final value with an MPS bond dimension ($M$) of only about 200. This is to be contrasted with the extremely slow convergence of DMRG algorithm with both localized and delocalized $k-$space basis. Given that the cpu time of both the MPS-LCC and the DMRG algorithm scale as $O(M^3)$ we see about three orders of magnitude improvement in the computational cost.

Now we assess the performance of the \emph{projector approximation}, where the zeroth order wavefunction is only approximately evaluated. Here again the perturbation theory shows rapid convergence towards the FCI energy. In particular, when the energy of $\Psi^{(0)}$ is minimized by using an MPS with a virtual bond dimension of only 20 the errors in energy compared to full MPS-LCC rapidly diminishes as shown in Table~\ref{tab:hub}. 

Two $p-$MPS-LCC calculations, one with an (8e, 8o) and other with (24e, 24o) active space were performed on a 2-D Hubbard model with 50 sites. The zeroth order wavefunction in the two cases only accounted for about $1\%$ and $14\%$ of correlation energy respectively. But remarkably, the third order correction to the energy was able to capture 95\% and 99\% of the remaining correlation energy. The first order correction to the wavefunction in the two cases above were represented by an MPS of bond dimension 5000 and 20000 respectively. Based on the results of the smaller 18 site Hubbard model we expect this perturbation theory to be rapidly convergent although it was not possible to carry out these calculations due to the high computational cost.

\begin{table*}
\caption{\label{tab:hub}The table shows the results of a MPS-LCC and $p-$MPS-LCC calculations (Energy/electron in units of $t$) to various orders of perturbation theory. For the 50 site Hubbard model we have performed two MPS-LCC calculations, one with an active space of 8 electrons in 8 orbitals ($p-$MPS-LCC(8)) and the other with 24 electrons in 24 orbitals ($p-$MPS-LCC(24)). We have also shown the coupled cluster results up to the CCSDTQP level, calculated using the program \textsc{MRCC}\cite{Rolik2013,kallay} for the 18 site Hubbard model (the CC calculations didn't converge for the 50 site Hubbard model).}
\begin{tabular}{lcccccc}
\hline
\hline
Order&\multicolumn{3}{c}{18 Site 2D Hubbard}&&\multicolumn{2}{c}{50 Site 2D Hubbard}\\
\cline{2-4}\cline{6-7}
of Theory&MPS-LCC&$p-$MPS-LCC&CC&&$p-$MPS-LCC(8)&$p-$MPS-LCC(24)\\
\cline{1-4}\cline{6-7}
0&	-0.804 & -0.802&-0.778& &-0.679&-0.705\\	
2&	-0.949	&-0.948&	-0.959&&-0.873&-0.867\\
3&	-0.960	&-0.961&	-0.965&&-0.867&-0.878\\
4&	-0.959	&-0.960&	-0.958&&$-$&$-$\\
5&	-0.958	&-0.959&	-0.958&&$-$&$-$\\
6&	-0.958	&-0.958&	$-$&&$-$&$-$\\
7&	-0.958	&-0.958&$-$	&&$-$&$-$\\	
\cline{1-4}\cline{6-7}
\bf{FCI}& \multicolumn{3}{c}{\bf{-0.958}$^a$}&&\multicolumn{2}{c}{\bf{-0.880}$^b$}\\
\hline
\end{tabular}
\flushleft{\footnotesize{$^a$ Obtained by exact diagonalization, $^b$Auxiliary-field Monte Carlo\cite{PhysRevE82}, which has no sign problem at half filling.}}
\end{table*}

\subsubsection*{3-band Hubbard model} 
\begin{table}
\caption{Table shows the calculated ground state energy for the 3-band Hubbard model. Here $E_0$ is the zeroth order energy obtained by fully correlating 10 holes in the 10 lowest energy orbitals. It is remarkable that the relatively much cheaper MPS-LCC theory is accurate to 4 significant places compared to the expensive DMRG calculation.}\label{tab:3band} 
\begin{tabular}{cccc}
\hline
FCIQMC&$E_0$&MPS-LCC&DMRG\\
\hline
-1.5817(5)&1.3399&1.5821&1.5819\\
\hline
\end{tabular}
\end{table}
Recently Schwarz et al.\cite{PhysRevB91} have published FCIQMC\cite{booth:054106,cleland:041103} results on an undoped 3-band ($p-d$) Hubbard model with 10 unit cells. Each unit cell containing CuO$_2$ is represented by three orbitals, one 3$d_{x^2-y^2}$ centered on the Cu atom and a $2p_x$ and $2p_y$ orbital on the the Oxygen atoms displaced in the $x-$direction and $y-$directions relative to the Cu respectively. For the details of the Hamiltonian we refer the reader to the original publication, but we would like to note that the model has inter-site potential and is extremely strongly correlated with the on-site repulsion divided by nearest neighbor hopping $U/t \approx 8$. We performed an MPS-LCC calculation, where 10 holes  were exactly correlated in the 10 lowest energy orbitals and the effect of the remaining 20 orbitals was taken into account perturbatively with a first order wavefunction represented by an MPS of bond dimension $M=1500$. Table~\ref{tab:3band} shows that the resulting MPS-LCC energy has an astonishingly small error of less than 0.0002 eV/hole compared to a very large DMRG calculation with an $M=6000$.\\

\subsection*{Conclusion and Outlook}
In this paper we show that the governing equations of multireference LCC theory can be efficiently solved using MPS by slightly modifying the DMRG algorithm. The theory has been used to obtain highly accurate energies, with a fraction of the cost of a variational DMRG calculation, of several benchmark problems: \emph{ab-initio} Hamiltonian of diamond, to Cr$_2$ dimer, two-dimensional 1-band and 3-band Hubbard models at half filling in the strongly correlated regime of $U/t=4-8$. More broadly, our work demonstrates new possibilities for efficiently accessing the ground state wave functions of highly correlated materials like transition metal oxides fully \emph{ab-initio} without recourse to approximate models. 
 
\begin{acknowledgements}
The calculations made use of the facilities of the Max Planck Society's Rechenzentrum Garching.
\end{acknowledgements}


\end{document}